\documentclass[prb,preprint]{revtex4-1} 

\usepackage{amsmath}  
\usepackage{amsfonts} 
\usepackage{graphicx} 
\usepackage{epstopdf}
\usepackage{esint}

\begin{document}


\title{A Pedagogical Model of Static Friction}

\author{Galen T. Pickett}
\email{Galen.Pickett@csulb.edu} 
\affiliation{Department of Physics and Astronomy, California State University Long Beach, 1240 Bellflower Blvd., Long Beach, CA 90840, USA} 


\date{\today}

\begin{abstract}
While dry Coulombic friction is an elementary topic in any standard introductory course in mechanics, the critical distinction between the kinetic and static
friction forces is something that is both hard to teach and to learn.
In this paper, I describe a geometric model of static friction that may help 
introductory students to both understand and apply the Coulomb static
friction approximation.
\end{abstract}

\maketitle 

\section{Introduction}
One of the key disputes that had to be settled when the modern formulation
of mechanics was being constructed was whether friction had a primary or
secondary role.
A hard-won step forward was made with 
the recognition  
that the inertia principle (momentum in a system can only be changed through
an interaction with its surroundings) is fundamental, and not the ``friction principle'' (that systems naturally and continuously slow down unless acted upon by their
surroundings).
That the ``friction principle'' seems to be so naturally confirmed by everyday
experience is one of the big challenges in teaching the inertia principle, and
is the source of many naive pre- and mis-conceptions on the part of students.

So, it is no small victory when a student builds sufficient mastery with Newtonian
ideas to successfully solve the usual spectrum of problems
set  in a first-semester 
mechanics course.  The pedagogical problem I wish to explore here is a common
one in these courses (both at the high school and university level). 
The Coulomb model of dry friction is simple enough to state, and while 
the Coulombic kinetic friction is fraught with its own set of pitfalls for students 
\cite{teach,change}, the {\em static} Coulomb friction approximation is (in my
experience) even more tangled with students' conceptual difficulties.

Indeed, the static Coulomb friction approximation puts an upper limit on a quantity 
that students have to work extraordinarily hard to convince themselves 
{\em even exists}.
This paper is (yet one more \cite{least_force,static_measure,tatic_measure}) attempt to give students a sense of what this
maximum force means, and why the angle of repose is a good measure 
of the static coefficient of friction, and to put this pedagogical tool into the
hands of instructors.  

\section{Elementary Problem}
\begin{figure}[h!]
\centering
\includegraphics[scale=0.5]{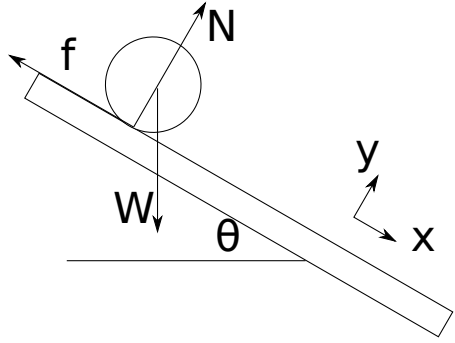}
\caption{A disc of moment of inertia $I$ rolls without slipping on a 
surface with static coeficient of friction $\mu_s$.}
\label{figure1}
\end{figure}

An elementary problem will illustrate the conceptual problem that I encounter 
all too often in my own first-semester mechanics course.  The situation is
shown in Figure~\ref{figure1}.
A uniform cylinder of radius R is placed on a ramp inclined from the horizontal at an
angle $\theta$.  The coefficient of static friction and kinetic friction, 
$\mu_s$ and $\mu_k$ are given.  Assuming that the object rolls without slipping,
{\it what is its acceleration down the ramp}?

Taking the $x$-axis as going down the ramp, and the $y$ axis as normal to the
ramp. a good student will generate the following from Newton's second law:
\begin{equation}
m g \sin \theta - f = m a,
\end{equation}
and
\begin{equation}
- m g \cos \theta + N = 0.
\end{equation}
The first is the resolution of the weight and friction force in the $x$-direction
as being the explanation for the $x$-component of the acceleration of the
disc, and the second is a really quite sophisticated statement that the 
constraint normal force $N$ will take {\em whatever value is required} so that
the observed motion ($a_y=0$) is maintained.  
At this point, there are three unknowns in the problem, $f, a$ and $N$, and it
is all too tempting to close the set of equations with the statement of Coulomb kinetic friction - because the disc is {\em moving}:
\begin{equation}
f = \mu_k N. \label{bad}
\end{equation}
This is, of course, an excellent solution to a completely different problem, not the 
one at hand.  For, the resulting (incorrect)
acceleration is just that of an object sliding
down a incline:
\begin{equation}
a= g (\sin \theta - \mu_k \cos \theta).
\end{equation}
But, this is completely unsatisfactory, and I attempt to guide students to
discover why this is a bad solution from a skeptical review of the result.
For instance, we could easily choose $\tan \theta = \mu_k$ so that the
``acceleration'' would vanish.  Thus, at this ``magic'' angle, a disc placed at rest on such a surface would remain in equilibrium - defying every experience every student has had with objects rolling down hill.  We can make $a$ small, but it
is very hard to imagine how it can be made to vanish at finite $\theta$ ... or
even become negative!

The resolution is trivial for an instructor, of course.  We have not even used the
``rolling without slipping'' constraint, and the critical role the moment of
inertia of the disc plays in the physics has not been established.
In fact, the friction force is determined by the rolling without slipping constraint:
if $f$ were too big (small), the disc would over (under) rotate, and the point of contact between the disc and the ramp would slide.
The correct analysis eventually leads to
\begin{equation}
a = g \sin \theta \frac{m R^2}{I + m R^2},
\end{equation}
where $I$ is the moment of inertia of the disc and 
\begin{equation}
\tan \theta < \mu_s,
\end{equation}
ensuring that the friction force stays below its maximal value.

And, while this is indeed a trivial problem that we expect any well-trained student to be able to solve, there is still the slightly (to me) unsatisfactory 
and qualitative difference between the kinetic friction force (which has a
given magnitude proportional to the loading of the surfaces) and the
static friction force (which has the character of a constraint force) with
its mysterious maximal value.  
As a student, the idea that the kinetic friction force was the result of
asperities on each surface interacting was convincing and suggestive.
What, however, is being ``broken'' in the static friction force?  Dry welding 
of asperities, or adhesion of the surfaces, or any other number of 
phenomenon are called to mind as at least a metaphor for Coulomb 
dry static friction.  
While dry friction is far too complex a phenomenon to explain with a simple
geometric theory, I can explain its general features in a pedagogical and 
geometric toy model.

\section{$\mu_s$ as a Geometric Parameter}
\begin{figure}[h!]
\centering
\includegraphics[scale=0.5]{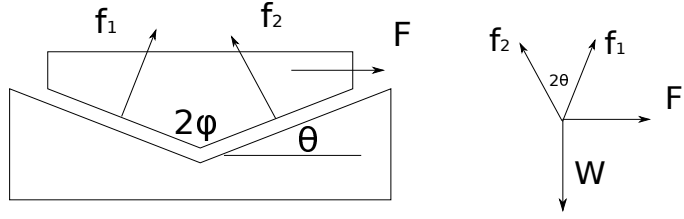}
\caption{A wedge of half-opening angle $\phi$ is resting on a conforming trough with tangent parameter $\theta = \pi/2-\phi$.  The net force on the wedge consists of two reaction forces,
$f_1$ and $f_2$, the weight $W$ and the driving force, $F$.  A free body
diagram is shown at right for the wedge.}
\label{figure2}
\end{figure}

Before anyone mucks around with modeling surface interactions with
friction, we introduce the physicist's toy model of the ``frictionless'' surface,
and we train students quite well in how to analyze systems built from
these imaginary surfaces.
Consider the symmetric wedge of mass $m$ and bottom angle $2 \phi$ resting 
on a conforming frictionless surface, as in Figure~\ref{figure2}.

I place a loading horizontal force $F$ on the wedge and ask two simple questions: what are the two forces
$f_1$ and $f_2$, and is there a maximum force $F$ that I can apply and
still maintain equilibrium?
Clearly, $f_1$ and $f_2$ are ``ordinary'' normal forces, and can only point 
in the given direction, but with an unknown magnitude.
Equilibrium requires
\begin{equation}
(f_1-f_2) \cos \phi  - F =0
\end{equation}
and
\begin{equation}
(f_1 + f_2) \sin \phi - m g = 0.
\end{equation}
Thus, the values of $f_1$ and $f_2$ can be found, although this is not
such an illuminating result:
\begin{equation}
f_1 = \frac{1}{2} \left( \csc \phi \mbox{  } m g - \sec \phi \mbox{  }F\right),
\end{equation}
and
\begin{equation}
f_2= \frac{1}{2} \left( \csc \phi \mbox{  }m g + \sec \phi \mbox{  } F\right).
\end{equation}
Not surprisingly, $f_2$ has a larger magnitude than $f_1$, because $F$ is
dragging the wedge off to the right.
Much more enlightening is to consider the {\em maximal value} of $F$ consistent
with equilibrium.
Here, because the left surface is providing $f_1$ by {\em pressing away} from itself, the maximal value of $F$ will occur when $f_1=0$:
\begin{equation}
F = \cot \phi \mbox{  } mg = \tan \theta \mbox{  } mg.
\end{equation}
Here, we have replaced the half-opening angle of the wedge, $\phi$ with the
incline anlge of the surface $\theta = \pi/2 - \phi$.
Thus, if the lateral force $F$ satisfies:
\begin{equation}
F < \mu_s N \mbox{  with  } \mu_s = \tan \theta  \mbox{   and   } N = m g,
\end{equation}
then the wedge will remain in equilibrium.  Should $F$ exceed this value, 
the wedge will start to accelerate.
This is a simple, straightforward model for Coulomb static friction.

\section{Discussion}
\begin{figure}[h!]
\centering
\includegraphics[scale=0.5]{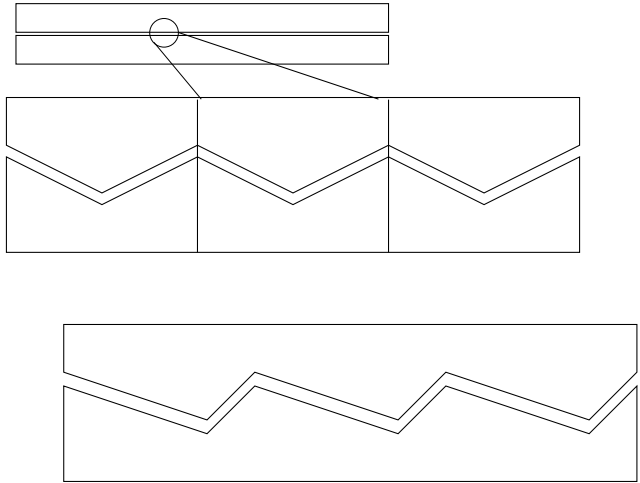}
\caption{Two macroscopic flat surfaces modeled locally as two sawtooth frictionless surfaces in contact.  Manifestly, the surfaces will begin to slide past each other when the tilt angle of the planks exceeds the roughness angle of the surface.  The lower panel shows the possibility of an asymmetry between a ``back'' and ``forward'' static friction force.}
\label{figure3}
\end{figure}

Very well, a single wedge resting in a conforming trough behaves similarly
to an object resting on a flat surface characterized by static friction $\mu_s$.
If we conceptualize such a flat, frictioned surface as consisting of two surfaces as in Figure~\ref{figure3} we have a useful toy model for static friction.
Here, the upper and lower surfaces are frictionless in the usual sense, but consist of a uniform sawtooth profile with a ``roughness'' angle $\theta$.
If a lateral force $F$ is applied to the upper object, the argument above implies
that each wedge-shaped subregion will remain in equilibrium only so long as
\begin{equation}
F < \tan \theta n_o m_o g \equiv \tan \theta N_{tot},
\end{equation}
where $n_o$ is the total number of wedges in the upper surface, each consisting of a mass $m_o$.

If we were to tilt the lower surface at an angle $\alpha$, as long as
$\alpha < \theta$, there will be a nonzero reaction force between the two
surfaces, and the object will remain motionless.  When $\alpha > \theta$ the frictionless surfaces making up the system are free to slide past each other.
Thus, the angle of repose matches the usual Euler condition:
\begin{equation}
\tan \alpha = \mu_s \equiv \tan \theta.
\end{equation}

This toy model could be made very explicit by adding low-friction caster wheels to the wedge and trough in Figure~2, and could be used as a lecture demonstration or as a stand alone experiment in equilibrium leading to a study of static friction.

The thing that I enjoy so much about this model is that it places kinetic and static friction upon the same footing.  That the kinetic coefficient of friction between two surfaces is related (in some hard to analyze but definite manner) to the {\em shapes} or the surfaces, as characterized by a  RMS tangent angle 
$\left<\theta\right>$ is a fixture in elementary mechanics texts.  
This model makes it clear that static friction arises from the same basic phsyics - two rough surfaces interacting.

An interesting, implication of the model is that it may be possible to create surfaces with anisotropic friction,\cite{anisotropic,cnt} but in more than just the sense of their being low friction axes on a surface.  Indeed, if microscopically corrugated surfaces with a different forward roughness, $\theta_{f} \ne$ the backward roughness $\theta_b$, different threshold forces in the forward and reverse directions could be engineered.  
Thus, not just an axis but a specific direction on a surface - {\em a tribological
diode} could be engineered.

\section{Conclusion}
I have constructed a simple mechanical toy model for dry static Coulombic friction, and naturally interpreted the roughness of a surface as the source of
all of the important static friction properties introduced in a first-semester 
mechanics course.  
Surfaces are modeled as frictionless but impenetrable, and it is the tangent angle of the sawtooth contacts surfaces make with each other that gives rise to $\mu_s$.

\end{document}